\documentclass{article}
\usepackage{amssymb}
\def\a{{\alpha}}
\def\b{{\beta}}
\def\c{{\gamma}}

\def\nn{ \nonumber }
\def\bq{ \begin{equation} }
\def\eq{ \end{equation} }
\def\ben{ \begin{eqnarray} }
\def\en{ \end{eqnarray} }
\def\frac#1#2{{#1\over #2}}
\def\dfrac#1#2{{\displaystyle{#1\over#2}}}

\begin{document}
%%%%%%%%%%%% TITLE   %%%%%%%%%%%%%%
\title{On the Drach superintegrable systems.}
\author{
 A.V. Tsiganov\\
{\small\it
 Department of Mathematical and Computational Physics,
 Institute of Physics,}\\
{\small\it St.Petersburg University, 198 904,  St.Petersburg, Russia,}\\
{\small\it e-mail: tsiganov@mph.phys.spbu.ru} }
\date{}
\maketitle

%%%%%%%%  A B S T R A C T  %%%%%%%%%
\vskip0.5cm {\small
Cubic invariants for two-dimensional degenerate Hamiltonian systems are
considered by using variables of separation of the associated St\"ackel
problems with quadratic integrals of motion. For the superintegrable
St\"ackel systems the cubic invariant is shown to admit new
algebro-geometric representation that is far more elementary than the
all the known representations in physical variables. A complete list of
all known systems on the plane which admit a cubic invariant is
discussed.
\vskip1truecm}

%%%%%%%%%%%%  TEXT  %%%%%%%%%%%%%%
\section{Introduction}
\setcounter{equation}{0}
In 1935 Jules Drach applied direct method for search of the integrable
Hamiltonian systems of two degrees of freedom, which admit a cubic
second integral \cite{dr35}. Recall, the direct approach leads to a
complicated set of nonlinear equations, whose nonlinearity has no {\it a
priory} restriction. We can attempt to solve these second invariant
differential equations using various simplifying assumptions. The Drach
ansatz for the Hamilton function $H$ and for the second cubic invariant
$K$
\ben
H&=&p_xp_y+ U(x,y)\,,\nn\\
\label{drint}\\
K&=&6w(x,y)\,\left(\dfrac{\partial H}{\partial x}\,p_y
-p_x\,\dfrac{\partial H}{\partial y}\right) -P(p_x,p_y,x,y)\nn
\en
yields ten new integrable systems. Recall, for the the fixed polynomial
$P(p_x,p_y,x,y)$ the potential $U(x,y)$ and function $w(x,y)$  be
solution of some differential equations \cite{dr35}.

For the other natural Hamilton functions
\bq
H=p_x^2+p_y^2+V(x,y)
\eq
the similar approach has been used by Fokas and Lagerstrom \cite{fl80},
by Holt \cite{ho82} and by Thompson \cite{thomp84}. Note, that the some
Drach results have been rediscovered in these papers.

The most complete classifications of known results was later brought
together by Hietarinta in 1987 \cite{hiet87}. Recently \cite{kr98},
two-dimensional hamiltonian systems with the cubic integrals were
investigated using the Jacobi change of the time. In \cite{kr98} a
complete list of all known systems was extended in comparison with
\cite{hiet87}.

The aim of this note is to study the Drach systems and some other
degenerate systems on the plane with the cubic in momenta integrals of
motion. We prove that eight Drach hamiltonians belong to the Stackel
family of integrals \cite{st95} and, moreover, seven of them are
degenerate systems.

Recall, the system is called superintegrable or degenerate if the
Hamilton function $H$ is in the involution with two integrals of motion
$I$ and $K$, such that
\bq
\{H,I\}=\{H,K\}=0\,,\qquad \{I,K\}=J(H,I,K)\,.
\label{jint}
\eq
Initial integrals and  the constant of motion $J(H,I,K)$ are generators
of the polynomial associative algebra \cite{bdk94,vin95}, whose defining
relations are polynomials of certain order in generators.

Below we shall consider two-dimensional systems with two quadratic
integrals of motion $I_1=H,~I_2=I$ and one qubic integral $K$. So, by
the Bernard-Darboux theorem \cite{w73} the system with integrals
$I_1,~I_2$ belong to the St\"ackel family of integrable systems
\cite{st95}. Therefore, let us begin with remaining of
some necessary results about the St\"{a}ckel systems \cite{st95}.

\section{The St\"{a}ckel systems}
\setcounter{equation}{0}
The systems associated with the name of St\"{a}ckel \cite{st95} are
holonomic systems on the phase space ${\mathbb R}^{2n}$ equipped with
the canonical variables $\{p_j,q_j\}_{j=1}^n$. The nondegenerate
$n\times n$ St\"{a}ckel matrix $\bf S$, with entries $s_{kj}$ depending
only on $q_j$
\[
\det {\bf S}\neq 0\,,\qquad \dfrac{\partial s_{kj}}{\partial q_m}=0\,,
\quad j\neq m\]
defines $n$ functionally independent integrals of motion
\bq
I_k=\sum_{j=1}^n c_{jk}\left(p_j^2+U_j\right)\,,
\qquad c_{jk}=\dfrac{{\mathfrak s}^{kj}}{\det {\bf S}}\,,
\label{int}
\eq
which are quadratic in momenta. Here ${\bf C}=[c_{jk}]$ denotes inverse
matrix to $\bf S$ and  ${\mathfrak s}^{kj}$ be cofactor of the element
$s_{kj}$. The common level surface of these integrals
\[M_\a=\left\{z\in {\mathbb R}^{2n}: I_k(z)=\a_k\,,~k=1,\ldots,n\right\}\]
is diffeomorphic to the $n$-dimensional real torus and one immediately
gets
\bq
p_j^2=\left(\dfrac{\partial {\cal S}}{\partial q_j}\right)^2=
\sum_{i=1}^n \a_i s_{ij}(q_j)-U_j(q_j)
\,.
\label{curvj}
\eq
Here ${\cal S}(q_1\,\ldots,q_n)$ is a reduced action function
\cite{st95}. For the rational
entries of $\bf S$ and  rational potentials $U_j(q_j)$ one gets
\bq
p_j^2=\dfrac{\prod_{i=1}^k(q_j-e_i)}{\varphi^2_j(q_j)}
\,,
\label{curvn}
\eq
where $e_i$ are constants of motion and functions $\varphi_j(q_j)$
depend on coordinate $q_j$ and numerical constants
\cite{ts97d}. The Riemann surfaces (\ref{curvn}) are isomorphic to the canonical
hyperelliptic curves
\bq
{\cal C}_j:\quad
\mu_j^2=\prod_{i=1}^k(\lambda-e_i)\,,\qquad \mu_j=\varphi(q_j)\,p_j\,,
\label{sthc}
\eq
where the senior degree $k$ of polynomial fixes the genus
$g_j=[(k-1)/2]$ of the algebraic curve ${\cal C}_j$. Considered
together, these curves determine an $n$-dimensional Lagrangian
submanifold in ${\mathbb R}^{2n}$
\[
{\cal C}^{(n)}:\qquad {\cal C}_1(p_1,q_1)\times{\cal C}_2(p_2,q_2)
\times\cdots\times{\cal C}_n(p_n,q_n)\,.\]
The Abel transformation linearizes equations of motion on ${\cal
C}^{(n)}$ by using first kind abelian differentials on the corresponding
algebraic curves \cite{ts98b}. The basis of first kind abelian
differentials is uniquely related to the St\"ackel matrix $\bf S$
\cite{ts97d,ts98b}.

Now let us turn to the superintegrable or degenerate systems in the
classical mechanics. One of the main examples of the two-dimensional
superintegrable systems is the isotropic harmonic oscillator, which has
many common properties with the Drach degenerate systems.  Recall, for
the oscillator the Hamilton function and the second integral of motion
look like
\[ H=p_1^2+p_2^2+q_1^2+q_2^2\,,\qquad I=p_1^2+q_1^2 - p_2^2 -q_2^2\,.
\]
Obviously, the angular momentum
\bq
K=q_1p_2-p_1q_2=\dfrac12\left( p_1\,\dfrac{d p_2}{dt} -
\dfrac{dp_1}{dt}\,p_2\right)\,.
\label{kint}
\eq
is the third integral of motion. Two pairs of quadratic integrals
$I_1=H,~I_2=I$ and $\tilde{I}_1=H,~\tilde{I}_2=K^2$ are associated with
the following St\"ackel matrices
\[
{\bf S}=\left(\begin{array}{cc}
  1 & 1 \\
  1 & -1
\end{array}\right)\,,\quad\mbox{\rm and}\quad
\tilde{\bf S}=\left(\begin{array}{cc}
  1 & 0 \\
  r^{-2} & 1
\end{array}\right)\,,\quad r^2=x^2+y^2
\]
respectively. So, the corresponding equations of motion may be separated
in the different curvilinear coordinate systems.

For this degenerate St\"ackel system and for all other known
superintegrable St\"{a}ckel systems with quad\-ra\-tic integrals of
motion the number degree of freedom $n>g$ is always more then sum of
genuses $g_j$ of the corresponding algebraic curves. In this case the
number of independent first kind abelian differentials be insufficient
for the inversion of the Abel-Jacobi map on ${\cal C}^{(n)}$.

To construct inversion of this map for the degenerate systems one has to
complete a given basis of the differentials to the set of $n$
differentials. We have some freedom in a choice of complimentary
differentials and, therefore, we can associate the different St\"ackel
matrices to one given Hamilton function \cite{ts98b}. By using first
kind abelian differentials one gets superintegrable St\"ackel system
with quadratic integrals only. Of course, we can try to add the second
and third kind abelian differentials, but we do not know such examples.

Below we prove that for all the known  superintegrable systems with a
qubic integral $K$ the number degree of freedom $n=2$ is more then sum
$g=g_1+g_2=0$ of genuses $g_j=0$ of the associated Riemann surfaces
(\ref{sthc}) too. The corresponding dynamics is splitting on two spheres
\bq
{\cal C}_{1,2}:\qquad \mu^2=\a_{1,2}\lambda^2+\b_{1,2}\lambda
+\c_{1,2}\,,\qquad g_{1,2}=0\,,
\label{tori}
\eq
where $\a_j,\b_j$ and $\c_j$ be the constants of motion.

In variables $\mu_{1,2}$ (\ref{sthc}) the additional cubic integral of
motion for all the degenerate Drach systems looks like
\bq
K=\dfrac{\det {\bf S}}{s_{21}s_{22}}\,
\left(\mu_1\,\dfrac{d \mu_2}{dt}-\dfrac{d \mu_1}{dt}\,\mu_2\right)\,.
\label{3int}
\eq
This generalized "angular momentum" gives rise to the first order
integrals (\ref{kint}) or the third order polynomials in momenta
depending on the St\"ackel matrices $\bf S$ and potentials $U_j$. In our
case it will be qubic integral, which coincides with the initial Drach
integral up to numerical factor.

To consider nonlinear algebra of integrals of motion for the Drach
systems we shall introduce new generators $\{N,a,a^\dag\}$ instead of
the two quadratic integrals $I_1=H,~ I_2=I$, one qubic integrals $K$ and
the constant of motion $J$ (\ref{jint}). Similar to oscillator these new
generators have the following properties
\bq
\begin{array}{ll}
  \{N,a\}=a\,, &  \{N,a^\dag\}=-a^\dag\,, \\
  \\
  \{a,a^\dag\}=\Phi(I_1,I_2)\,,\qquad &  a\,a^\dag=\Psi(I_1,I_2)\,.
\end{array}
\label{osc}
\eq
Here generator $N(I_1,I_2)$, functions $\Psi(I_1,I_2)$ and
$\Phi(I_1,I_2)$ depend on the quadratic St\"ackel integrals only. Two
other generators $a$ and $a^\dag$ be functions on the all three
constants of motion $I_1,I_2,K$, such that
\[K=\rho(I_1,I_2)\,(a-a^\dag)\,,\qquad J(I_1,I_2,K)=\dfrac{a+a^\dag}2\,.
\]
The relations (\ref{osc}) remind the deformed oscillator algebra, which
is widely used for the superintegrable systems with quadratic integrals
of motion \cite{bdk94,vin95}. However, instead of the usual quadratic
algebra of integrals we shall get more complicated algebras of
integrals.

\section{The Drach systems}
\setcounter{equation}{0}
Let us reproduce corrected Drach results in his notations
\ben
(a)\qquad U&=&\dfrac{\a}{xy}+\b x^{r_1}y^{r_2}+\c
x^{r_2}y^{r_1}\,,\quad\mbox{\rm where}\quad r_j^2+3r_j+3=0\,,
\label{a}\\
\nn\\
\qquad P&=&({xp_x}-{p_y}y)^3\,,\qquad w= x^2y^2/2\,,\nn \\
\nn\\
(b)\qquad U&=&\dfrac{\a}{\sqrt{xy}}+\dfrac{\b}{(y-\mu
x)^2}+\dfrac{\c\,(y+\mu x)}{\sqrt{xy}\,(y-\mu x)^2}\,,\label{b}\\
\nn\\
\qquad P&=&3(xp_x-p_yy)^2(p_x+\mu p_y)\,,
\qquad w=xy(y-\mu x)\,,\nn\\
\nn\\
(c)\qquad
U&=&\a\,xy+\dfrac{\b}{(y-ax)^2}+\dfrac{\c}{(y+ax)^2}\,,\label{c}\\
\nn\\
\qquad P&=&3(xp_x-p_yy)(p_x^2-a^2p_y^2)\,,
\qquad w=(y^2-a^2x^2)/2\,,\nn\\
\nn\\
(d)\qquad U&=&\dfrac{\a}{\sqrt{y(x-a)\,}}+\dfrac{\b}{\sqrt{y(x+a)\,}}
+\dfrac{\c x}{\sqrt{x^2-a^2\,}}\,,\label{d}\\
\nn\\
\qquad P&=&3p_y\,\left[(xp_x-p_yy)^2-a^2p_x^2\right]\,,
\qquad w=-y(x^2-a^2)\,,\nn\\
\nn\\
(e)\qquad U&=&\dfrac{\a}{\sqrt{xy\,}}+\dfrac{\b}{\sqrt{x\,}}
+\dfrac{\c}{\sqrt{y\,}}\,,\label{e}\\
\nn\\
\qquad P&=&3p_yp_x\,(xp_x-p_yy)\,,
\qquad w=-2xy\,,\nn\\
\nn\\
(f)\qquad U&=&\a\,xy +\b y\dfrac{2x^2+c}{\sqrt{x^2+c\,}}+\dfrac{\c
x}{\sqrt{x^2+c\,}}\,,\label{f}\\
\nn\\
\qquad P&=&3p_y^2\,(xp_x-yp_y)\,,\qquad w=(x^2+c)/2\,,\nn
\nn\\
(g)\qquad
U&=&\dfrac{\a}{(y+3mx)^2}+\b(y-3mx)+\c(y-mx)(y-9mx)\,,\label{g}\\
\nn\\
\qquad P&=&(p_x+3m p_y)^2\,(p_x -3mp_y)\,,
\qquad w=-m(y+3mx)\,,\nn\\
\nn\\
(h)\qquad U&=&(y+\dfrac{mx}3)^{-2/3}\,\left[\a+\b\,(y-mx/3)+
\c\,(y^2-\dfrac{14mxy}3+\dfrac{m^2x^2}9)\right]\,,\label{h}\\
\nn\\
\qquad P&=&(p_x-\dfrac{mp_y}3)
\left(p_x^2+\dfrac{10mp_xp_y}{3}+\dfrac{m^2p_y^2}9\right)\,,
\qquad w=-m(y+\dfrac{mx}3)\,,\nn\\
\nn\\
(k)\qquad U&=&\a y^{-1/2} +\b x y^{-1/2} +\c x\,,\label{k}\\
\nn\\
\qquad P&=&3p_x^2p_y\,,\qquad w=-y\,,\nn\\
\nn\\
(l)\qquad U&=&\a\left(y-\dfrac{\rho x}3\right)+\b x^{-1/2} +\c
x^{-1/2}(y-\rho x)\,,\label{l}\\
\nn\\
\qquad P&=&3p_xp_y^2+\rho p_y^3\,,\qquad w=x\,.\nn
\en
Here $\a$, $\b$, $\c$, $\mu$, $\rho$, $a$, $c$, and $m$  be arbitrary
parameters. In compare with \cite{dr35} we corrected function $w$ in the
case (g) (\ref{g}) and revised potential $U$ in the case (k) (\ref{k}).
Namely this corrected Hamiltonian is in the involution with the initial
Drach cubic integral $K$ (\ref{drint}).

With an exception of three cases (a) (\ref{a}), (h) (\ref{h}) and (l)
(\ref{l}), other Drach systems are degenerate or superintegrable
St\"{a}ckel systems. The separation variables associated with the pair
of quadratic integrals $\{I_1=H,I_2\}$ are the St\"ackel variables.
Equations of motion may be integrated in quadratures \cite{ts97d}, but
these quadratures depend on the value of quadratic integral $I_2$. Thus,
instead of the solution of initial Drach problem related to integrals
$\{H,K\}$ we shall solve the associated problem with quadratic integrals
$\{H,I_2\}$.

In the case (h) (\ref{h}) we also have the St\"{a}ckel systems
\cite{ts98c}.  Only in this case
(h) (\ref{h}) dynamics is splitting on two tori and the number degrees
of freedom is equal to the sum of genuses $g=n=2$, such that the
corresponding system is non-degenerate.

In the case (l) (\ref{l}) the Hamilton function coincides with the
hamiltonian of the previous St\"ackel system (\ref{k}) at $\rho=0$. Here
we shall not consider this generalized St\"ackel system at $\rho\neq 0$.

Below we shall consider the Drach integrals (\ref{drint}) up to linear
transformations of the coordinates and a rescaling of these integrals.
It allows us to remove some parameters in the Hamilton functions without
loss of generality. To associate the degenerate Drach hamiltonians with
the St\"ackel matrices $\bf S$ we can join these systems into the four
pairs of the systems with a common matrices $\bf S$.

\subsection{Case (a)}
In our previous paper \cite{ts99c}, the first Drach system (\ref{a}) has
been related to the three-particle periodic Toda lattice in the
center-of-mass frame. Namely, after canonical change of the time
$t=q_{n+1}$ and the Hamiltonian $H=p_{n+1}$ at the extended phase space
\[ d\widetilde{t}=(xy)^{-1}\cdot dt\,,
 \qquad \widetilde{H}=xy\cdot\,(H+\delta)\,,\]
and after further canonical transformation of other variables
\[
x=e^{\frac{q_1+iq_2}2}\,,\quad
p_x=(p_1-ip_2)e^{-\frac{q_1+iq_2}2}\,,\qquad
y=e^{\frac{q_1-iq_2}2}\,,\quad p_y=(p_1+ip_2)e^{-\frac{q_1-iq_2}2}\,,
\]
the Hamilton function (\ref{a}) becomes
\[
\widetilde{H}=p_1^2+p_2^2+\beta e^{\displaystyle
-\frac{1}2q_1-\frac{\sqrt{3}}2q_2} +\gamma e^{\displaystyle
-\frac{1}2q_1+\frac{\sqrt{3}}2q_2} +\delta e^{\displaystyle q_1}+\a\,.
\]
It is the Hamiltonian of the tree-particle periodic Toda lattice in the
center-of-mass frame. The separation variables survive at the change of
the time. Thus, for the first Drach system we can separate variables and
integrate equations of motions in quadratures repeating the calculations
for the Toda chain \cite{ts99d}.

Later in  \cite{thomp84} Thompson considered this system too. In fact,
after point transformation
\[x=r\,e^{i\phi}\,,
\quad p_x=\dfrac{e^{-i\phi}}2\left(p_r-ip_\phi\,r^{-1}\right)\,,
\qquad
y=r\,e^{-i\phi}\,,\quad
p_y=\dfrac{e^{i\phi}}2\left(p_r+ip_\phi\,r^{-1}\right)\,
\]
the Drach hamiltonian $H$ (\ref{drint}) looks like
\[
H=p_r^2+\dfrac{p_\phi^2}{r^2}+U(r,\phi)\,,
\]
up to numerical factor. Namely this Hamilton function was studied in
\cite{thomp84} and \cite{kr98}. The special substitution of the potential
$U(r,\phi)$ into the Drach equations leads to the following equation
\[
U(r,\phi)=\dfrac{f(\phi)+f''(\phi)}{r^3}\,,\quad
\Rightarrow
\quad f'''\,f''-2f''\,f'-3f'f=0\,,\]
introduced in \cite{kr98}. Of course, the same equation follows from the
functional equation on the Toda potential \cite{hiet87}.

\subsection{Cases (b) and (e)}
Put $\mu=1$ in (\ref{b}). Let us introduce the St\"ackel matrix
\bq
{\bf S}_{be}=\left(\begin{array}{cc} q_1^2 &q_2^2  \\
\\ 1 & 1\end{array}\right)\,,
\label{msbe}
\eq
and take the following potentials
\[
\begin{array}{lll}
  (b)\qquad &U_1=2\a-\dfrac{\b-2\c}{q_1^2},
     \qquad &U_2=-2\a-\dfrac{\b+2\c}{q_2^2}, \\
  \\
  (e)\qquad &U_1=2\a+2(\b+\c)q_1,
     \qquad &U_2=-2\a-2(\b-\c)q_2.
\end{array}
\]
The corresponding Hamilton functions $I_1$ (\ref{int}) coincide with the
Hamilton functions $H$ for the Drach systems (\ref{b}) and (\ref{e}),
after the following canonical point transformation
\[
x=\dfrac{(q_1-q_2)^2}4, \quad p_x=\dfrac{p_1-p_2}{q_1-q_2},
\qquad
y=\dfrac{(q_1+q_2)^2}4, \quad p_y=\dfrac{p_1+p_2}{q_1+q_2}.
\]
The second integrals of motion $I_2$ (\ref{int}) are second order
polynomials in momenta. The third independent integrals of motion $K$
are defined by  (\ref{3int}), where
\[(b)\quad \mu_1=q_1p_1,\quad \mu_2=q_2p_2,\qquad (e)\quad
\mu_1=p_1,\quad \mu_2=p_2\,.
\]
From the above definitions we can introduce generators of the nonlinear
algebra of integrals (\ref{osc}) and  verify properties of this algebra
\ben
(b)\qquad && N=\dfrac{I_2}{4\sqrt{H}}\,,
\qquad a=J+4\sqrt{H}\,K\,,\qquad
       a^\dag=J-4\sqrt{H}\,K\,,\nn\\
\nn\\
\qquad &&a\,a^\dag=16\left(4H\,(\b+2\c)-(2\a+I_2)^2\right)
                     \left(4H\,(\b-2\c)-(2\a-I_2)^2\right)\,,\nn\\
\nn\\
\qquad &&\{a,a^\dag\}=-256\sqrt{H}\,
\left(I_2\,(I_2-2\a)(I_2+2\a)-4H\,(\b I_2-4\a\c)\right)\,,\nn
\en
and
\ben
(e)\qquad &&N=\dfrac{I_2}{2\sqrt{H}}\,,
\qquad a=J+2\sqrt{H}\,K\,,\qquad
       a^\dag=J-2\sqrt{H}\,K\,,\nn\\
\nn\\
\qquad &&a\,a^\dag=-16\left(H\,(2\a+I_2)-(\b-\c)^2\right)
                      \left(H\,(2\a-I_2)+(\b+\c)^2\right)\,,\nn\\
                      \nn\\
\qquad && \{a,a^\dag\}=-64H^{3/2}\,(I_2\,H-\b^2-\c^2)\,.\nn
\en

\subsection{Cases (c) and (g)}
Put $a=1$ in (\ref{c}) and  $m=1/3$ in (\ref{g}). Let us introduce the
St\"ackel matrix
\bq
{\bf S}_{cg}=\left(\begin{array}{cc} \dfrac12 &-\dfrac12  \\
\\ 1 & 1\end{array}\right)\,,
\label{mscg}
\eq
and take the following potentials
\bq
\begin{array}{lcc}
  (c)\qquad &U_1= \dfrac{\a q_1^2}4+\dfrac{\c}{q_1^2},
     \qquad &U_2=\dfrac{\a q_2}4-\dfrac{\b}{q_2^2}, \\
  \\
  (g)\qquad &U_1=-\dfrac{\c q_1^2}3+\dfrac{\a}{q_1^2},
     \qquad &U_2=-\dfrac{4\c q_2^2}3-\b q_2.
\end{array}
\label{ucg}
\eq
The corresponding Hamilton functions $I_1$ (\ref{int}) coincide with the
Hamilton functions $H$ (\ref{c}) and (\ref{g}), after the following
canonical point transformation
\[
x=\dfrac{q_1-q_2}2,\quad p_x=p_1-p_2,
\qquad y=\dfrac{q_1+q_2}2, \quad
p_y=p_1+p_2.
\]
The second integrals of motion $I_2$ (\ref{int}) are the second order
polynomials in momenta. The third independent integrals $K$ are defined
by (\ref{3int}), where
\[
(c)\quad \mu_1=q_1p_1\quad\mu_2=q_2p_2\,,
\qquad
(g)\quad \mu_1=q_1p_1\,,\quad\mu_2=p_2\,.
\]
Generators and defining relations of the nonlinear algebra of integrals
(\ref{osc}) look like
\ben
(c)
\qquad &&N=\dfrac{I_2}{2\sqrt{-\a}}\,,
\qquad a=J+2\sqrt{-\a}\,K\,,\qquad
       a^\dag=J-2\sqrt{-\a}\,K\,,\nn\\
\nn\\
\qquad &&a\,a^\dag=\left(H^2+4H\,I_2+4I_2^2-4\a\c\right)
                   \left(H^2-4H\,I_2+4I_2^2+4\a\b\right)\,,\nn\\
                      \nn\\
\qquad && \{a,a^\dag\}=-32\sqrt{a}\,
\left(I_2\,(H-2I_2\,)(H+2I_2\,)-\a\b(H+2I_2\,)-\a\c(H-2I_2\,)
\right)\,,\nn
\en
and
\ben
(g)
\qquad &&N=\dfrac{I_2}{4}\sqrt{\dfrac3{\c}}\,,
\qquad a=J+4\sqrt{\dfrac{\c}3}\,K\,,\qquad
  a^\dag=J-4\sqrt{\dfrac{\c}3}\,K\,,\nn\\
\nn\\
\qquad &&a\,a^\dag=1/9\left(8\c H-16\c I_2+3\b^2\right)
                      \left(3H^2+12H\,I_2+12I_2^2+16\a\c\right)\,,\nn\\
                      \nn\\
\qquad && \{a,a^\dag\}=64\left(\dfrac{\c}3\right)^{3/2}\,\left(
\dfrac{(2I_2+H)(4\c H-24\c I_2+3b^2)}{4\c}-\dfrac{16\a\c}{3}
\right)\,.
\nn
\en

\subsection{Cases (d) and (f)}
Put $a=1$ in (\ref{d}) and $c=1$ in (\ref{f}). Let us introduce two the
St\"ackel matrices
\bq
{\bf S}_d==\left(\begin{array}{cc}
 1 &1 \\
 \\ \dfrac1{q_1^2} & \dfrac1{q_2^2}\end{array}\right)\,,\qquad
{\bf S}_f=\left(\begin{array}{cc}
 \dfrac1{q_1} &\dfrac1{q_2} \\
 \\ \dfrac1{q_1^2} & \dfrac1{q_2^2}\end{array}\right)
\label{msdf}
\eq
and takes the following potentials
\[
\begin{array}{lcc}
(d)\qquad   &U_1= 2\c +\dfrac{2\sqrt{2}(\a+\b)}{q_1},
     \qquad &U_2=-2\c -\dfrac{2\sqrt{2}(\a-\b)}{q_2},\\
  \\
(f)  \qquad &U_1=\dfrac{\c}{2q_1}+\dfrac{(\a+2\b)}4,
     \qquad &U_2=\dfrac{\c}{2q_2}+\dfrac{(\a-2\b)}4.
\end{array}
\]
The corresponding Hamilton functions $I_1$ (\ref{int}) coincide with the
Hamilton functions $H$ (\ref{d}) and (\ref{f}) up to numerical factor,
after the following explicit canonical transformations
\ben
(d)\quad &x=\dfrac{q_1^2+q_2^2}{2q_1q_2},
\quad p_x=\dfrac{(p_1q_1-p_2q_2)q_1q_2}{q_1^2-q_2^2},\qquad
&y=q_1q_2, \quad p_y=\dfrac{p_1q_1+p_2q_2}{2q_1q_2},\nn\\
\nn\\
(f)\quad &x=\dfrac{q_1-q_2}{2\sqrt{q_1q_2}},
\quad p_x=\dfrac{2(p_1q_1-p_2q_2)\sqrt{q_1q_2}}{q_1+q_2},\qquad
&y=\sqrt{q_1q_2}, \quad p_y=\dfrac{p_1q_1+p_2q_2}{\sqrt{q_1q_2}}.\nn
\en
The second integrals of motion $I_2$ (\ref{int}) are the quadratic
polynomials in momenta. The third independent integrals $K$ are defined
by (\ref{3int}), where for the both systems one gets
\[\mu_1=q_1p_1\,,\qquad\mu_2=q_2p_2\,.\]
Generators of the nonlinear algebra of integrals (\ref{osc}) are given
by
\[
N=\sqrt{I_2}\,,
\qquad a=J+2\sqrt{I_2\,}\,K\,,\qquad
       a^\dag=J-2\sqrt{I_2\,}\,K\,,
\]
which have the following properties
\ben
(d)
\qquad &&a\,a^\dag={16}
\left(I_2\,(2\c-H)+2(\a+\b)^2\right)
\left(I_2\,(2\c-H)-2(\a-\b)^2\right)\,,\nn\\
\nn\\
\qquad &&\{a,a^\dag\}=64\sqrt{I_2\,}
\left(I_2\,(2\c-H)(2\c+H)+2(\a^2+\b^2) H+8\a\b\c\right)
\,,\nn\\
\nn\\
(f)\qquad &&a\,a^\dag=\frac1{16}
\left(4I_2\,(\a+2\b)+(\c-2H)^2\right)
\left(4I_2\,(\a-2\b)+(\c+2H)^2\right)\,,\nn\\
\nn\\
\qquad &&\{a,a^\dag\}=-4\sqrt{I_2\,}
\left((2\b-\a)(2\b+\a)I_2+\a H^2+2\b\c H+\dfrac{\a\c^2}4\right)
\,.\nn
\en

\subsection{Cases (h) and (k)}
Put $m=3$ in (\ref{h}). Let us introduce two the St\"ackel matrices
\bq
{\bf S}_h==\left(\begin{array}{cc}
   q_1 &q_2  \\
 \\ -1 & 1\end{array}\right)\,,\qquad
{\bf S}_k=\left(\begin{array}{cc}
 q_1&-q_2\\
 \\ 1 & 1\end{array}\right)
\label{mshk}
\eq
and take the following potentials
\[
\begin{array}{lcc}
(h)\qquad&U_1=\dfrac{\a\c-\dfrac{\b^2}{16}}{4\c}-\dfrac{2\c\,q_1^3}9,
   \qquad&U_2=\dfrac{\a\c-\dfrac{\b^2}{16}}{4\c}-\dfrac{2\c\,q_2^3}9\\
  \\
(k)  \qquad &U_1= {\a}+\b q_1+\dfrac{\c q_1^2}2,
     \qquad &U_2=-{\a}+\b q_2+\dfrac{\c q_2^2}2.
\end{array}
\]
The corresponding Hamilton functions $I_1$ (\ref{int}) coincide with the
Hamilton functions $H$ (\ref{h}) and (\ref{k}) up to numerical factor,
after the following explicit canonical transformations
\[
\begin{array}{lcc}(h)\qquad
&x=\dfrac{p_2-p_1}{4\sqrt{\c}} +
\dfrac{(3q_1+3q_2)^{3/2}}{54}+\dfrac{\b}{16\c},\qquad
&p_x=3\dfrac{p_1+p_2}{\sqrt{3q_1+3q_2\,}}+
\sqrt{\c}(q_1-q_2),
\\
\\
\\
& y= -\dfrac{p_2-p_1}{4\sqrt{\c}} +
\dfrac{(3q_1+3q_2)^{3/2}}{54}-\dfrac{\b}{16\c},
\qquad
& p_y=3\dfrac{p_1+p_2}{\sqrt{3q_1+3q_2\,}}- \sqrt{\c}(q_1-q_2),
\\
\\ \mbox{\rm and}\qquad&&\\
\\
(k)\qquad& x=\dfrac{q_1-q_2}2,\quad p_x=p_1-p_2,
&y=\dfrac{(q_1+q_2)^2}4,
\quad p_y=\dfrac{p_1+p_2}{q_1+q_2}\,.
\end{array}
\]
Note, in the case (h) (\ref{h}) we used non-point canonical
transformation in contrast with other Drach systems.

In the last case (k) (\ref{k}) integral of motion $I_2$ (\ref{int}) is
the second order polynomial in momenta. The third independent integral
$K$ is defined by (\ref{3int}), where
\[\mu_1=p_1\,\quad\mbox{\rm and}\quad \mu_2=p_2.
\]
Generators and defining relations of nonlinear algebra of integrals
(\ref{osc}) look like
\ben
(k)\qquad && N=\dfrac{I_2}{\sqrt{-2\c}}\,,
\qquad a=J+\sqrt{-2\c}\,K\,,\qquad
       a^\dag=J-\sqrt{-2\c}\,K\,,\nn\\
\nn\\
\qquad&& a\,a^\dag=\left(2\c\,(I_2+\a)+(H+\b)^2\right)
                     \left(2\c\,(I_2-\a)+(H-\b)^2\right)\,,\nn\\
\nn\\
\qquad&& \{a,a^\dag\}=-4\c\sqrt{-2c}\,\left(H^2+2\c I_2+\b^2\right)\,.\nn
\en

In the case (h) (\ref{h}) the second integral of motion $I_2$
(\ref{int}) is the second order polynomial in momenta $\{p_1,p_2\}$.
However, after the non-point transformation of variables this integral
$I_2$ becomes the qubic in momenta $\{p_x,p_y\}$ Drach integral $K$
(\ref{h}). The corresponding dynamics is splitting on two tori and the
third order polynomial (\ref{3int}) does not commute with the Hamilton
function. Later this system has been rediscovered by Holt \cite{ho82}.

\section{Other degenerate systems on the plane with a qubic integral of motion}
\setcounter{equation}{0}
In this section we consider the St\"ackel systems on the plane with a
qubic integral of motion defined by the following  Hamilton function
\[H=\dfrac12\Bigl(p_x^2+p_y^2\Bigr)+V(x,y)\,.
\]
As above, the corresponding qubic integral will be written at the Drach
form (\ref{drint}).

On the plane we know four orthogonal systems of coordinates: elliptic,
parabolic, polar and cartesian. Thus, we reproduce all the known results
\cite{hiet87,kr98} in correspondence with the type of the
associated St\"ackel matrix \cite{ts97d,ts98b}.

The systems whose Hamilton functions separable in cartesian coordinates:
\ben
%34
(A)\qquad V&=&\a\,(4 x^2+y^2)+\b\,x + \dfrac{\c}{y^2}\,,\label{A}\\
\nn\\
\qquad P&=&p_x\,p_y^2\,,\qquad w=\dfrac{y}6\,,\\
\nn\\
%6
(B)\qquad V&=&\a\,(x^2 + y^2)+\dfrac{\b}{x^2}+\dfrac {\c}{y^2}
\,,\label{B}\\
\nn\\
\qquad P&=&( x\,p_y - y\,p_x)\,p_x\,p_y\,,\qquad w=\dfrac{xy}6\,,\nn\\
\nn\\
%7
(C)\qquad V&=&\a\,(x^2 + y^2) +\b\,\dfrac{xy}{(x^2-y^2)^2}\,,\label{C}\\
\nn\\
\qquad P&=&(x\,p_y-y\,p_x)\,\left(p_x^2-p_y^2\right)\,,
\qquad w=\dfrac{x^2-y^2}6\,,\nn\\
\nn\\
%5
(D)\qquad V&=&\a(9x^2 + y^2)\,,\label{D}\\
\nn\\
\qquad P&=&(x\,p_y - p_x\,y)\,p_y^{2}\,, \qquad w=-\dfrac{y^2}{18}\,,
\nn
\en
The systems whose Hamilton functions separable in parabolic coordinates:
\ben
%8
(F)\qquad V&=&(\a+\dfrac{\b}{r+x} + \dfrac{\c}{r-x})\,r^{-1}\,,\quad
r=\sqrt{x^2+y^2}\,,\label{F}\\
\nn\\
\qquad P&=&(x\,p_y-p_x\,y)^2\,p_x\,,\qquad w=\dfrac{y\,r^2}{12}\,,\nn\\
\nn\\
%10
(G)\qquad V&=&(\a+\dfrac{\b\,x}{y^2})\,r^{-1}\,,\label{G}\\
\nn\\
\qquad P&=&(x\,p_y-p_x\,y)^2\,p_x\,,\qquad w=\dfrac{y\,r^2}{12}\,,\nn\\
\nn\\
%11
(H)\qquad V&=& (\a+\b\,\sqrt{r+x}+\c\,\sqrt{r-x})\,r^{-1}\,,\label{H}\\
\nn\\
\qquad P&=&(x\,p_y-p_x\,y)\,\left(2p_x^2+2p_y^2-
\dfrac{\b}{\sqrt{r+x}}-\dfrac{\c}{\sqrt{r-x}}\right)\,,
\qquad w=-\dfrac{r^2}6\,,\nn
\en
One system with the Hamilton function separable in polar coordinates:
\ben
(I)\qquad V&=&\a+\dfrac{\b}{\sqrt{x^2+y^2}}+\dfrac{\rho}{(\delta\,
x+\c\, y)^2}+
\dfrac{\c\,x-\delta\,y}{\sqrt{x^2+y^2}\,(\delta\,x+\c\,y)^2}\,,\label{I}\\
\nn\\
P&=&(p_x\,y-p_y\,x)(\c\,p_x-\delta\,p_y)\,,\qquad
w=\dfrac{(x^2+y^2)\,(\delta\,x+\c\,y)}{12}\,.\nn
\en
It is new superintegrable system with a cubic integral of motion, which
is  a deformation of the degenerate kepler model. An application of the
direct method \cite{fl80,ho82,hiet87} or the Jacobi method \cite{kr98}
does not allows us to obtain this system (\ref{I}).

Three exceptional systems whose qubic integral of motion $K$ we can not
rewrite in the "generalized angular momentum" form (\ref{3int}):
\ben
%1
(K)\qquad V&=&\a\,\sqrt{x}\,\pm\,\b\,\sqrt{y}\,,\label{K}
\qquad P=\dfrac{\b}{\a}\,p_x^3\, \mp\, \dfrac{\a}{\b}\,p_y^3\,,\qquad
w=\sqrt{xy}\,,\\
\nn\\
(L)\qquad V&=&\a\left(\sqrt{x} + \b\,y\right)\,,\label{L}
\qquad P=p_x^3\,,\qquad w=-\dfrac{\sqrt{x}}{2\b}\,,\\
\nn\\
(M)\qquad V&=&f'(\phi)\,r^{-2}\,,\label{M}
\qquad
K=p_\phi^2\,\left(\cos\phi\,p_r-\sin\phi\,r^{-1}\,p_\phi\right)+\\
\nn\\
&+&\left(2f'(\phi)\cos\phi-f(\phi)\sin(\phi)\right)\,p_r
+\left(3f'(\phi)\sin\phi+f(\phi)\cos\phi\right)r^{-1}p_\phi\,.\nn
\en
At the case (M) (\ref{M}) we used the standard polar coordinates
$\{r,p_r,\phi,p_\phi\}$ and the function $f(\phi)$ has to satisfy the
following equation
\[ f''\,\left(3f'\sin\phi+f\cos\phi\right)
+2f'\,\left(2f'\cos\phi-f\sin\phi\right)=0\,.
\]
At these exceptional cases the Hamilton functions
(\ref{K},\ref{L},\ref{M}) are separable at the cartesian and polar
coordinates, respectively. However, the St\"ackel potentials $U_{1,2}$
are not polynomials in variables of separation.

\subsection{Cartesian coordinates, cases A-D}
Let us introduce the St\"ackel matrix
\bq
{\bf S}_{A-D}=\left(\begin{array}{cc} \dfrac12 &\dfrac12  \\
\\ 1 & -1\end{array}\right)\,,
\label{msAD}
\eq
and take the following potentials
\bq
\begin{array}{lcc}
  (A)\qquad &U_1=8\a\,q_1^2+2\b\,q_1,
     \qquad &U_2=2\a\,q_2^2+\dfrac{2\c}{q_2^2}, \\
  \\
  (B)\qquad &U_1= 2\a\,q_1^2+\dfrac{2\b}{q_1^2},
     \qquad &U_2= 2\a\,q_2^2+\dfrac{2\c}{q_2^2},\\
   \\
  (C)\qquad &U_1=\dfrac{\a\,q_1^2}2-\dfrac{\b}{4q_1^2},
     \qquad &U_2=\dfrac{\a\,q_2^2}2+\dfrac{\b}{4q_2^2},\\
   \\
  (D)\qquad &U_1=18\a\,q_1^2,
     \qquad &U_2=2\a\,q_2^2.
\end{array}
\label{uAD}
\eq
The corresponding Hamilton functions $I_1$ (\ref{int}) coincide with the
Hamilton functions $H$ (\ref{A},\ref{B},\ref{D}) and  (\ref{C}) if
\[(A-B,D)\qquad  x=q_1,\,\qquad y=q_2\]
or after the following canonical transformation
\[
(C)\qquad x=\dfrac{q_1-q_2}2,\quad p_x=p_1-p_2,
\qquad y=\dfrac{q_1+q_2}2, \quad
p_y=p_1+p_2.
\]
The second integrals of motion $I_2$ (\ref{int}) are the second order
polynomials in momenta. The third independent integrals $K$ are
calculated by (\ref{3int}), where variables
\[
(A)\quad \mu_1=p_1\quad\mu_2=q_2p_2\,,
\qquad
(B-C)\quad \mu_1=q_1p_1\,,\quad\mu_2=p_2\,.
\]
determine the left hand side of the canonical algebraic curves
(\ref{sthc}). At the case (D) the variables
\[(D)\qquad \mu_1=p_2q_1-\dfrac{p_1q_2}3\,,\qquad
\mu_2=p_2q_2 \]
have not such natural algebro-geometric meaning.

Generators and defining relations of the nonlinear algebra of integrals
(\ref{osc}) look like
\[
(A-B)
\qquad N=\dfrac{I_2}{4\sqrt{-2\a}}\,,
\qquad a=J+4\sqrt{-2\a}\,K\,,\qquad
       a^\dag=J-4\sqrt{-2\a}\,K\,,\]
such that
\ben
(A)
\qquad a\,a^\dag&=&4(4\a(2I_2+H)+\b^2)\Bigl((2I_2-H)^2-64\a\c\Bigr)\,,\nn\\
                      \nn\\
\qquad  \{a,a^\dag\}&=&-
128\a\sqrt{-2\a}\,(2I_2-H)(6I_2+H+\dfrac{\b^2}{2\a})- 64\a\c)\,,\nn\\
\nn\\
(B)
\qquad a\,a^\dag&=&\Bigl((2I_2+H)^2-64\a\b\Bigr)\,
                   \Bigl((2I_2-H)^2-64\a\c\Bigr)\,,\nn\\
                      \nn\\
\qquad
\{a,a^\dag\}&=&-16\sqrt{-2\a}\,
\Bigl(\bigl((2I_2-H)^2-64\a\c\bigr)(2I_2+H)\nn\\
&&\qquad\qquad
    + \bigl((2I_2+H)^2-64\a\b\bigr)(2I_2-H)\Bigr) \,.\nn
\en
At the two last cases we have
\ben
(C)
\qquad N&=&\dfrac{I_2}{2\sqrt{-2\a}}\,,
\qquad a=J+2\sqrt{-2\a}\,K\,,\qquad
  a^\dag=J-2\sqrt{-2\a}\,K\,,\nn\\
\nn\\
\qquad a\,a^\dag&=&\Bigl((2I_2-H)^2-2\a\b\Bigr)
                   \Bigl((2I_2+H)^2+2\a\b\Bigr)\,,\nn\\
                      \nn\\
\qquad  \{a,a^\dag\}&=&-8\sqrt{-2\a}
   \Bigl(\bigl((2I_2-H)^2-2\a\b\bigl)(2I_2+H)\nn\\
&&\qquad\qquad+\bigl((2I_2+H)^2+2\a\b\bigr)(2I_2-H)\Bigr)\,,
\nn
\en
and
\ben
(D)
\qquad N&=&\dfrac{I_2}{6\sqrt{-2\a}}\,,
\qquad a=J+6\sqrt{-2\a}\,K\,,\qquad
  a^\dag=J-6\sqrt{-2\a}\,K\,,\nn\\
\nn\\
\qquad a\,a^\dag&=&-4(2I_2-H)^3\,(2I_2+H)\,,\nn\\
                      \nn\\
\qquad  \{a,a^\dag\}&=&
-96\sqrt{-2\a}\,(2I_2-H)^2\,(4I_2+H)
\,.
\nn
\en
At the case (D) (\ref{D}) the quantum counterpart of this qubic deformed
oscillator algebra has been used to study of the corresponding quantum
superintegrable system \cite{bdk94}.

\subsection{Parabolic coordinates, cases F-H}
Let us introduce two St\"ackel matrices
\bq
{\bf S}_{F,G}=\left(\begin{array}{cc} 1 &1  \\
\\q_1^{-1} & q_2^{-1}\end{array}\right)\,,\qquad
{\bf S}_{H}=\left(\begin{array}{cc} q_1^2 &q_2^2  \\
\\1 & 1\end{array}\right)\,,
\label{msFH}
\eq
and take the following potentials
\bq
\begin{array}{lcc}
  (F)\qquad &U_1=-\dfrac{\a}{2q_1}-\dfrac{\b}{2q_1^2},
     \qquad &U_2= \dfrac{\a}{2q_2}-\dfrac{\c}{2q_2^2}, \\
  \\
  (G)\qquad &U_1= \dfrac{\a}{2q_1}-\dfrac{\b}{4q_1^2},
     \qquad &U_2=-\dfrac{\a}{2q_2}+\dfrac{\b}{4q_2^2} ,\\
   \\
  (H)\qquad &U_1=-4\a - 8\sqrt{2}\b\,q_1,
     \qquad &U_2= 4\a+ 8\sqrt{-2}\c\,q_2.
   \end{array}
\label{uFH}
\eq
The corresponding Hamilton functions $I_1$ (\ref{int}) coincide with the
Hamilton functions $H$ (\ref{A},\ref{B}) and (\ref{C}) after the
following canonical point transformations
\[
(F,G)\qquad x=q_1+q_2\,,\quad p_x=\dfrac{p_1q_2-p_2q_2}{q_1-q_2}\,,\quad
y=2\sqrt{-q_1q_2}\,,\quad
p_y=\dfrac{(p_1-p_2)\sqrt{-q_1q_2}}{q_1-q_2}\,,
\]
and
\[(H)\qquad x=q_1^2+q_2^2\,,
\quad p_x=\dfrac12\dfrac{p_1q_1-p_2q_2}{q_1^2-q_2^2}\,,
\quad y=2iq_1q_2\,,\quad
p_y=\dfrac{i}2\dfrac{p_1q_2-p_2q_1}{q_1^2-q_2^2}\,.
\]
 The second integrals of motion $I_2$ (\ref{int}) are the second
order polynomials in momenta. The third independent integrals $K$ are
calculated by (\ref{3int}), where variables
\[
(F,G)\quad \mu_1=q_1p_1\quad\mu_2=q_2p_2\,,
\qquad
(H)\quad \mu_1=p_1\,,\quad\mu_2=p_2\,,
\]
define the left hand side of the canonical algebraic curves
(\ref{sthc}).

At all these cases (F,G) and (H) the generators and defining relations
of the nonlinear algebra of integrals (\ref{osc}) look like
\[
(F-H)
\qquad N=\dfrac{I_2}{2\sqrt{H}}\,,
\qquad a=J+2\sqrt{H}\,K\,,\qquad
       a^\dag=J-2\sqrt{H}\,K\,,\]
such that
\ben
(F)
\qquad a\,a^\dag &=&\dfrac{1}{16}\,
\Bigl(8\b\,H-(2I_2+\a)^2\Bigr)\,\Bigl(8\c\,H-(2I_2-\a)^2\Bigr)\,,\nn\\
\nn\\
\qquad  \{a,a^\dag\}&=&-2\sqrt{H}\,
\left(I_2(2I_2+\a)(2I_2-\a)-2H\Bigl(\b(2I_2-\a)+\c(2I_2+\a)
\Bigr)
\right)\,,\nn\\
\nn\\
(G)
\qquad a\,a^\dag&=&-\dfrac{1}{16}\,
\bigl(4\b\,H+(2I_2+\a)^2\bigr)
\,\bigl(4\b\,H-(2I_2-\a)^2\bigr)\,,\nn\\
                      \nn\\
\qquad
\{a,a^\dag\}&=&
2\sqrt{H}\,\bigl(I_2\,(2I_2+\a)(2I_2-\a)-2\a\b\,H\bigr) \,,\nn\\
\nn\\
(H)\qquad a\,a^\dag&=&16\,
\Bigl(H\,(I_2-4\a)+32\c^2\Bigr)
\,\Bigl(H\,(I_2+4\a)-32\b^2\Bigr)\,,\nn\\
                      \nn\\
\qquad
\{a,a^\dag\}&=&
-64{H}^{3/2}\,(H\,I_2-16\b^2+16\c^2) \,.\nn
\en

\subsection{Polar coordinates, case I}
Let us introduce St\"ackel matrix
\bq
{\bf S}_{I}=\left(\begin{array}{cc} 1 &0  \\
\\q_1^{-2} & 1\end{array}\right)\,,
\label{msI}
\eq
and take the following potentials
\bq
\begin{array}{lcc}
  (I)\qquad &U_1=\a+\dfrac{\b}{q_1},
     \qquad &U_2= \dfrac{\c\,\cos(q_2)-\delta\,\sin(q_2)+\rho}
                        {\Bigl(\delta\,\cos(q_2)+\c\,\sin(q_2)\Bigr)^2}, \\
     \end{array}
\label{uI}
\eq
The corresponding Hamilton function $I_1$ (\ref{int}) coincides with the
Hamilton function $H$ (\ref{I}) if $q_1=r$ and $q_2=\phi$ are the
standard polar coordinates on the plane. The second integrals of motion
$I_2$ (\ref{int}) are the second order polynomials in momenta. The third
independent integrals $K$ are calculated by (\ref{3int}), where
variables
\[
(I)\quad \mu_1=p_1\,,
   \qquad\mu_2=p_2\,\Bigl(\delta\,\cos(q_2)+\c\,\sin(q_2)\,\Bigr)\,,
\]
define the canonical algebraic curves (\ref{sthc}).

At $\delta=1$ and $\c=0$ the generators and defining relations of the
nonlinear algebra of integrals (\ref{osc}) look like
\ben
(I)
\qquad N&=&-{\sqrt{-I_2}}\,,
\qquad a=J+2\sqrt{-I_2}\,K\,,\qquad
       a^\dag=J-2\sqrt{-I_2}\,K\,,\nn\\
\nn\\
\qquad a\,a^\dag &=&(4\,H\,I_2-4\,\a\,I_2+\b^2)\,(4\,I_2^2-4\,\rho\,I_2+1)\,,\nn\\
\nn\\
\qquad  \{a,a^\dag\}&=&-8\,\sqrt{-I_2}\,
\Bigl((2\,I_2-\rho)\,\b^2+(12I_2^2-8\rho\,I_2+1)\,(H-\a)\Bigr)
\,.
\en
This system does not contain in the list of the known integrable systems
\cite{hiet87,kr98}. At the cases (I) (\ref{I}) and (M) (\ref{M})
we have common leading part $P$ of the qubic integrals $K$. However, at
the case (M) we can not rewrite the qubic integral in the "generalised
angular momentum" form.

\section{The Lax representation}
\setcounter{equation}{0}
In \cite{ts97d,ts98b} we proposed some construction of the $2\times 2$
Lax matrices for the St\"ackel systems with homogeneous St\"ackel
matrices \cite{ts98b} and with uniform potentials $U_j=U$. The
Drach-St\"ackel systems fall out from this subset of the St\"ackel
systems. Nevertheless, we could construct the $2\times 2$ Lax matrices
for these systems by using various covering \cite{ts97d} of the initial
spheres ${\cal C}_{1,2}$ (\ref{tori}).

Here we consider the $4\times 4$ Lax matrices for some Drach systems by
using canonical transformations of the extended phase space, which
induce transformations of the Lax matrices \cite{ts98b,ts99c}. Recall,
if the St\"{a}ckel matrices $\bf S$ and $\widetilde{\bf S}$ be
distinguished the first row only, the corresponding St\"{a}ckel systems
are related by canonical change of the time $q_{n+1}=t$ and conjugated
momenta $p_{n+1}=-H$
\cite{ts98b}
\bq
t\mapsto\widetilde{t}\,,\qquad d\widetilde{t}=
\dfrac{\det\widetilde{\bf S} }{\det{\bf S}}\,dt\,,
\qquad H\quad\mapsto\quad
\widetilde{H}=\dfrac{\det{\bf S}}{\det\widetilde{\bf S}}\,H\,.
\label{drttr}
\eq
Thus, starting with the St\"ackel systems related with matrix ${\bf
S}={\bf S}_{cg}$ (\ref{mscg}) we can study systems associated with
matrices $\widetilde{\bf S}={\bf S}_{be}$ (\ref{msbe}) and
$\widetilde{\bf S}={\bf S}_{k}$ (\ref{mshk}). Here subscripts mean the
type of the St\"ackel matrices for the different Drach systems.

The St\"ackel systems with the constant matrix ${\bf S}_{cg}$ possess
the following $4\times 4$ Lax matrices
\cite{van92,ts97d}
\bq
{\cal L}(\lambda)=
\left(\begin{array}{cc}
  L_1(\lambda,p_1,q_1) & 0 \\ \\
  0 & L_2(\lambda,p_2,q_2)
\end{array}\right)\,,
\label{drlax}
\eq
with independent $2\times 2$ non-trivial blocks $L_j(\lambda)$. For
instance, two standard blocks may be chosen
\[
L_j(\lambda)=\left(\begin{array}{cc}
  p_j & \lambda-q_j \\
  -\left[\dfrac{\phi_j}{\lambda-q_j}\right]_{MN} & -p_j
\end{array}\right)\,,\qquad
L_j(\lambda)=\left(\begin{array}{cc}
  \dfrac{p_jq_j}{\lambda} & 1-\dfrac{q_j^2}{\lambda} \\
  \dfrac{p_j^2}{\lambda}-\left[\dfrac{\phi_j}{1-\frac{q_j^2}{\lambda}}
  \right]_{MN} &
  -\dfrac{p_jq_j}{\lambda}
\end{array}\right)\,.
\]
Here $\phi(\lambda)$ is a parametric function on spectral parameter
$\lambda$ and $[\xi]_{N}$ is the linear combinations of the Laurent
projections \cite{ts97d}.

According to \cite{ts98b,ts99c}, canonical transformations of the
extended phase space induce shift of the Lax matrices depending on the
Hamilton function. Thus, by using one known Lax matrix ${\cal
L}(\lambda)$ (\ref{drlax}) we can construct another Lax matrices.
Namely, canonical transformations of the time (\ref{drttr}) give rise
the following shift of the corresponding Lax matrices
\bq
\widetilde{\cal L}(\lambda)={\cal L}(\lambda)-
\widetilde{H}\left(\begin{array}{cccc}
  0 & 0 & 0 & 0 \\
  a & 0 & 0 & 0 \\
  0 & 0 & 0 & 0 \\
  0 & 0 & b & 0
\end{array}\right)\,,\qquad a,b=\pm 1~\mbox{or}~\pm i\,,
\eq
where values of the constants $a,b$ depend on the chosen form of the
blocks $L_j(\lambda)$.

Below we present some Lax matrices constructed by designated above
scheme. In the case (c) the Lax matrix is given by
\[
{\cal L}_c(\lambda)=\left(\begin{array}{cccc}
  p_1 & \lambda-q_1 & 0 &0 \\
  (\lambda+q_1)\left(-\dfrac{\a}4+{\c}{q_1^2\lambda^2}\right) & -p_1 & 0 & 0 \\
  0 & 0 & ip_2 & i(\lambda-q_2) \\
  0 & 0 & i(\lambda+q_2)\left(-\dfrac{\a}4-{\b}{q_2^2\lambda^2}\right) & -ip_2
\end{array}\right)\,,
\]
so the spectral curve
\[\Gamma(\lambda,\mu):\qquad \det\left({\cal L}_c(\lambda)-\mu
I\right)=0\]
is a product
\[
\left(\mu^2-\dfrac{I_1}2-I_2+\dfrac{\a\lambda^2}4+\dfrac{\c}{\lambda^2}\right)
\left(\mu^2-\dfrac{I_1}2+I_2-\dfrac{\a\lambda^2}4+\dfrac{\b}{\lambda^2}\right)=0\,,
\]
of the corresponding canonical St\"ackel curves (\ref{tori}).

In the case (k) the Lax matrix is given by
\[
\widetilde{\cal L}_k=\left(\begin{array}{cccc}
  p_1 & \lambda-q_1 & 0 &0 \\
  -\dfrac{\c(\lambda+q_1)}2-\b & -p_1 & 0 & 0 \\
  0 & 0 & ip_2 & i(\lambda+q_2) \\
  0 & 0 & \dfrac{i\c(\lambda-q_2)}2+i\b & -ip_2
\end{array}\right)+
\widetilde{H}\left(\begin{array}{cccc}
  0 & 0 & 0 & 0 \\
  1 & 0 & 0 & 0 \\
  0 & 0 & 0 & 0 \\
  0 & 0 & i & 0
\end{array}\right)
\,,
\]
where $\widetilde{H}=I_1$ be the Hamilton function (\ref{k}). As in the
previous example the spectral curve
\[
\left(\mu^2-\dfrac{\c\lambda^2}2+(\b+I_1)\lambda+\a+I_2\right)
\left(\mu^2+\dfrac{\c\lambda^2}2+(\b-I_1)\lambda+\a-I_2\right)=0\,,
\]
is a product of the corresponding St\"ackel curves (\ref{tori}).

In the case (b) the Lax matrix is given by
\[
\widetilde{\cal L}_b=\left(\begin{array}{cccc}
  \dfrac{p_1q_1}{\lambda} & 1-\dfrac{q_1^2}{\lambda} & 0 &0 \\
  \\
  \dfrac{p_1^2-(\b-2\c)q_1^{-2}}{\lambda} & -\dfrac{p_1q_1}\lambda & 0 & 0 \\
  0 & 0 & \dfrac{p_2q_2}\lambda & 1+\dfrac{q_2^2}\lambda \\
  \\
  0 & 0 & \dfrac{-p_2^2+(\b+2\c)q_2^{-2}}\lambda & -\dfrac{p_2q_2}\lambda
\end{array}\right)+
\widetilde{H}\left(\begin{array}{cccc}
  0 & 0 & 0 & 0 \\
  1 & 0 & 0 & 0 \\
  0 & 0 & 0 & 0 \\
  0 & 0 & 1 & 0
\end{array}\right)
\,,
\]
where $\widetilde{H}=I_1$ be the Hamilton function (\ref{b}). The
corresponding spectral curve
\[\Gamma(y,\mu):\qquad
\det\left(\widetilde{\cal L}_b(\lambda)-y I\right)=0\]
is a product
\[
\left(y^2-I_1+\dfrac{2\a+I_2}\lambda-\dfrac{\b+2\c}{\lambda^2}\right)
\left(y^2-I_1+\dfrac{2\a-I_2}\lambda-\dfrac{\b-2\c}{\lambda^2}\right)=0\,,
\]
of the initial St\"ackel curves (\ref{curvn}), which could be rewritten
in the canonical form (\ref{tori}).

All the spectral curves of these $4\times 4$ Lax matrices $\cal L_c$,
$\widetilde{\cal L}_k$ and $\widetilde{\cal L}_b$ give rise to the
quadratic St\"ackel integrals $I_{1,2}$ (\ref{int}). The third integral
$K$ (\ref{3int}) may be extracted from the same matrices by using
multivariable universal enveloping algebras \cite{ts97d}. In fact, this
integral is a coefficient of the following multivariable polynomial
\bq
\mbox{\rm tr} P_{\pi}{\cal L}(\lambda_1)\otimes
{\cal L}(\lambda_2)\otimes {\cal L}(\lambda_3)\otimes {\cal
L}(\lambda_4)\,,
\label{256eq}
\eq
where $P_\pi$ be permutation operator of auxiliary spaces corresponding
to a  Young diagram $\pi$ \cite{ts97d}. The nonlinear algebra of
integrals (\ref{osc}) may be reproduced by using the Poisson bracket
relations between the Lax matrices $\otimes_j^k {\cal L}(\lambda_j)$
\cite{ts97d}. The formulae (\ref{256eq}) for the $256\times 256$
matrices has been proved by using the computer algebra system {\sl Maple
V}.

\section{Conclusion}
Let us discuss the list of all the known integrable natural Hamiltonian
systems in the plane with a qubic integral
\cite{dr35,hiet87,kr98}. We suppose that all these systems may be
embedded into the family of the St\"ackel systems \cite{ts98b}, either
into the subset of the generalized St\"ackel systems \cite{ts99c} or
these systems may be related to the Toda lattices and the Calogero-Moser
systems
\cite{hiet87,ts99c}. As an example the last case (l) (\ref{l}) of the
Drach systems and the Fokas-Lagerstrom \cite{fl80} model belong to the
generalized St\"ackel systems \cite{ts99c}. The complete classification
will be presented in the forthcoming publication.

In this note we proved this proposition for  all the Drach systems
\cite{dr35}. Moreover, we rewrite the qubic integrals
for the superintegrable Drach systems in common form (\ref{3int}). This
generalized "angular momentum" may be used to construct another
$n$-dimensional superintegrable St\"ackel systems with the cubic
integrals of motion.  For instance, let us consider the Hamilton
function
\bq
H=\dfrac12(p_x^2+p_y^2+p_z^2)+\dfrac{\c+\delta}{r} +\dfrac{1}{x^2+y^2}
\,
\left(\dfrac{\a\,(r-z)}{r}+\dfrac{\b(r+z)}{r}+U(y/x)\right)\,,
\label{3par}
\eq
where $r=\sqrt{x^2+y^2+z^2}$ and $U(y/x)$ be arbitrary function. The
corresponding equations of motion are separable in the parabolic
coordinates
\[q_1=r+z\,,\qquad q_2=r-z\,,\qquad q_3=\arctan(y/x)\,,\]
which related to the following St\"ackel matrix
\[\bf{S}=\left(\begin{array}{ccc}
  1 & 1 & 0 \\
  q_1^{-1} & q_2^{-1} & 0 \\
  q_1^{-2} & q_2^{-2} & -4
\end{array}\right)\,.
\]
The Hamilton function (\ref{3par}) coincides with the St\"ackel integral
$I_1$ (\ref{int}) if
\[U_1=\dfrac{\a}{q_1^2}+\dfrac{\c}{q_1}\,,\qquad
  U_2=\dfrac{\b}{q_2^2}+\dfrac{\delta}{q_2}\,,\qquad
  U_3=U(q_3)\,.\]
Thus we have integrable St\"ackel system with the independent integrals
of motion $I_{1,2}$ and $I_3$, which are quadratic polynomials in
momenta.

Canonical algebraic curves is defined in variables
\[\mu_1=p_1\,q_1\,,\qquad \mu_2=p_2\,q_2\,,\qquad \mu_3=p_3\,.\]
To substitute these variables in the  "generalized angular momentum"
(\ref{3int}) one gets additional qubic in momenta integral of motion
$K$. In the initial physical variables this integral $K$ looks like
\ben
K&=&(x^2+y^2)\,p_z^3 -2z\,p_z^2\,(p_x\,x+p_y\,y) -
p_z\,(p_x\,x+p_y\,y)^2\nn\\
\nn\\
&+& r^2
\left(\,
\dfrac{\partial H}{\partial z}\,(p_x\,x+p_y\,y)+
p_z\,\Bigr(\, p_x^2+p_y^2-x\dfrac{\partial H}{\partial
x}-y\dfrac{\partial H}{\partial y}\,\Bigl)\,\right)\,.\nn
\en
It will be interesting to understand the algebro-geometric origin of
this "generalized angular momentum" (\ref{3int}).

\end{document}